%% file: effcp.tex
\documentclass[11pt]{JHEP3}
\usepackage{amssymb}

\newcommand{\sv}{\mbox{\textsf{\textit{v}}}}
\newcommand{\sL}{\mbox{\textsf{\textit{L}}}}
\newcommand{\sR}{\mbox{\textsf{\textit{R}}}}
\newcommand{\VL}{V_{\rm L}}
\newcommand{\VR}{V_{\rm R}}
\newcommand{\YL}{Y_{\rm L}}
\newcommand{\YR}{Y_{\rm R}}
\newcommand{\mrl}{m_{\rm RL}}
\newcommand{\mlr}{m_{\rm LR}}
\newcommand{\PL}{P_{\rm L}}
\newcommand{\PR}{P_{\rm R}}
\newcommand{\UL}{U_{\rm L}}
\newcommand{\UR}{U_{\rm R}}
\newcommand{\Da}{\hat{D}}
\newcommand{\sm}{\mbox{\textsf{\textit{m}}}}
\newcommand{\sF}{\mbox{\textsf{\textit{F}}}}
\newcommand{\sD}{\mbox{\textsf{\textit{D}}}}
\newcommand{\sDa}{\hat{\sD}}
\newcommand{\m}[1]{{\underline{#1}}}

\input{newcomm}
\title{Effective CP violation in the Standard Model}

\author{Jan Smit\\
Institute for Theoretical Physics, University of Amsterdam, \\
       Valckenierstraat 65, 1018 XE Amsterdam, the Netherlands.\\
}


\preprint{ITFA-2004-27}

\abstract
{We study the strength of effective CP violation originating from the CKM
matrix in the effective action obtained by integrating out the
fermions in the Standard Model. Using results obtained by Salcedo
for the effective action in a general chiral gauge model, we find
that there are no CKM CP-violating terms to fourth order in a
gauge-covariant derivative expansion that is non-perturbative in
the Higgs field. The details of the calculation suggest that, at
zero temperature, the strength of CP violation is approximately independent of
the overall scale of the Yukawa couplings. Thus, order of
magnitude estimates based on Jarlskog's invariant could be too
small by a factor of about $10^{17}$.}

\begin{document}
\section{Introduction}
The weak, electromagnetic and strong
interactions described by the Standard Model (SM) have played a
role in the shaping of the universe as we know it
\cite{KT}. It is natural to assume that the same is true for finer
details, such as the CP violation embodied in the
Cabibbo-Kobayashi-Maskawa (CKM) matrix
\cite{Cabibbo:1963yz,Kobayashi:1973fv}. Particularly striking is
the fact that this CP violation, which is compatible with
experiment \cite{Eidelman}, can occur only with three or
more families, and three families are being observed. Yet, it is
often stated that the CP-violation caused by the CKM matrix is too
weak to be able to play a significant role in the generation of
the baryon asymmetry in the universe.

A scenario that has received considerable attention over the years
is electroweak baryogenesis \cite{Kuzmin:1985mm,Cohen:1993nk,Rubakov:1996vz},
in which the asymmetry is supposed to be generated during the
electroweak transition. One way to approach the problem of dealing
with complicated non-perturbative dynamics is to concentrate on
the bosonic variables by `integrating out the fermions'. CP
violation then enters the description effectively through
higher-dimensional terms in an effective lagrangian. The simplest of
these has been assumed to have the form
\cite{Shaposhnikov:1987tw,Shaposhnikov:1988pf}
\be
\frac{3\delta_{\rm CP}}{16\pi^2 M^2}\,
\vr^{\dagger}\vr\; \tr(A^{\mu\nu}\tilde{A}_{\mu\nu}),
\label{CPterm}
\ee
where $A_{\mu\nu}$ is the SU(2) field strength tensor,
$\tilde A_{\mu\nu}$ its dual,
$M$ is a mass depending on the scale of the problem and
$\dl_{\rm CP}$ is a dimensionless constant characterizing the strength
of the induced CP violation. In case of the finite-temperature
electroweak transition, a natural choice for $M$ is the
temperature $T$, and the usual estimate for $\dl_{\rm CP}$ is
given by
\cite{Jarlskog:1985ht,Shaposhnikov:1987tw,Shaposhnikov:1988pf}
\be
\dl_{\rm CP} = J\,
(m_{u}^{2}-m_{c}^{2}) (m_{c}^{2}-m_{t}^{2}) (m_{t}^{2}-m_{u}^{2})
(m_{d}^{2}-m_{s}^{2}) (m_{s}^{2}-m_{b}^{2}) (m_{b}^{2}-m_{d}^{2})
/T^{12} \approx 10^{-19},
\label{dlcpest1}
\ee
where $m_u$, \ldots,$m_b$ are the quark masses\footnote{We use
$m_u=0.0025$, $m_d=0.0045$, $m_s=0.09$, $m_c=1.26$, $m_b=4.26$, $m_t=175$ GeV.},
we used $T = 100$ GeV,
and \cite{Eidelman}
\be
J = |{\rm Im} (V_{fg}V_{hi}V_{fi}^* V_{hg}^*)| = (2.88 \pm 0.33)
\times 10^{-5}
\label{J}
\ee
is the simplest rephasing-invariant combination of the CKM matrix
$V$ \cite{BigiSanda,Branco:1999fs}.
Since the above estimate is many orders of magnitude smaller that
the baryon asymmetry $n_B/n_{\gm}\simeq 6 \times 10^{-10}$, the
usual conclusion is that CKM CP-violation cannot have been
instrumental in early universe baryogenesis.

Recently, new scenarios for electroweak baryogenesis have been put
forward in which the electroweak transition is supposed to have
been a tachyonic one at the end of inflation,
in which the effective squared Higgs mass
parameter turned negative in the early universe, {\em not} due to a
change in temperature but because of the coupling to a changing
inflaton field
\cite{Garcia-Bellido:1999sv,Krauss:1999ng,Garcia-Bellido:2003wd,Tranberg:2003gi}.
At the end of electroweak-scale inflation \cite{Copeland:2001qw,vanTent:2004rc}
the temperature is supposed to be zero, whereas the dynamics in
tachyonic transitions is dominated by the low-momentum modes of the fields
\cite{Skullerud:2003ki,Garcia-Bellido:2002aj}.
This suggests reconsidering the above order of
magnitude estimate for $\dl_{\rm CP}$
in an environment at zero temperature. Then the
quark masses in (\ref{dlcpest1}) are to be replaced by the Yukawa
couplings $\lm_u$, \ldots, $\lm_t$ ($\lm_u = \sqrt{2}\,m_u/v$,
etc., $v=246$ GeV), giving
\be
\dl_{\rm CP} = J\,
(\lm_{u}^{2}-\lm_{c}^{2}) (\lm_{c}^{2}-\lm_{t}^{2})
(\lm_{t}^{2}-\lm_{u}^{2}) (\lm_{d}^{2}-\lm_{s}^{2})
(\lm_{s}^{2}-\lm_{b}^{2}) (\lm_{b}^{2}-\lm_{d}^{2})
\approx 10^{-22},
\label{dlcpest2}
\ee
even smaller than (\ref{dlcpest1}). But what to use for $M$? A
natural choice is the (renormalized) expectation value of the Higgs field
$\langle\vr^{\dagger}\vr\rangle$.
In a low-temperature tachyonic electroweak transition
this increases from zero to close to its vacuum expectation value
$v^2/2$, suggesting a boost of the resulting CP violation when
$\langle\vr^{\dagger}\vr\rangle$ is small.

However, even with the Higgs field settled in its v.e.v., the
measured CP violating effects in accelerator experiments are at a
much higher level than $10^{-23}$ \cite{Eidelman}.
This suggests that the above order of magnitude estimates of
$\dl_{\rm CP}$ are misleading, at least at zero temperature
(see also \cite{Farrar:1993sp,Farrar:1994hn,Farrar:1994kf,Konstandin:2003dx}).

In this article we investigate CP-violation induced by the CKM
matrix using results for the effective bosonic action in a general
chiral gauge theory obtained by Salcedo \cite{Salcedo:2000hp,Salcedo:2000hx}. He presented
remarkably explicit results to fourth order in a gauge-covariant derivative
expansion, with coefficient-functions that are non-perturbative in the
Higgs field. Specializing these results to the case of the SM we found
that they do not contain CKM CP-violation (unfortunately).
However, the general form of the results suggests
strongly that the magnitude of the CP violation to be expected in
higher order is primarily set by the CKM-invariant $J$ in
(\ref{J}) and not by the tiny product of Yukawa couplings in
(\ref{dlcpest2}).

In section \ref{salc} we review the results of Salcedo that are
relevant for our purpose and apply these to the SM case in section
\ref{appSM}, in so far at they are relevant to CKM CP violation.
Considerations on the magnitude of CKM CP-violation
are in section \ref{magn}. In section \ref{gWZW} we show how the
CP-violating QCD $\theta$-term may be uncovered from the effective action
and our conclusions in section \ref{concl}.
In the appendix we give details some details of the functions
calculated by Salcedo.

\section{Salcedo's results}
\label{salc}
Salcedo calculated the fermion contribution to the euclidean
effective action for the Bose fields in a derivative expansion up
to fourth order in the gauge-covariant derivatives. The
effective action, $W$, corresponds to a model with $n$ Dirac fields and
is formally given  by
\be
W = -\Tr(\ln D),
\ee
where $D$ is a Dirac operator of the form
\bea
&&
D = D^{\rm R}_{\mu} \gm_{\mu} \PR + D^{\rm L}_{\mu} \gm_{\mu}\PL +
m_{\rm LR} \PR + m_{\rm RL} \PL,
\\
&&
\PR = \half(1+\gm_5),
\;\;\;\;
\PL = \half(1-\gm_5),
\eea
with\footnote{Our $\gm_4 = \gm_0({\rm Salcedo})$ and we made the
sign choice $\et_4({\rm Salcedo}) = +1$.}
$\gm_{\mu} = \gm_{\mu}^{\dagger}$ and
$\gm_5 = \gm_5^{\dagger} = - \gm_1\gm_2\gm_3\gm_4$. The covariant
derivatives $D^{\rm L,R}_{\mu} = \dmu + v^{\rm L,R}_{\mu}$ depend
on chiral $U(n) \times U(n)$ gauge fields $v^{\rm L,R}_{\mu}= -
(v^{\rm L,R}_{\mu})^{\dagger}$,
and $m_{\rm LR}$
and $m_{\rm RL}= (m_{\rm LR})^{\dagger}$ are $n\times n$ matrix
scalar fields transforming under gauge transformations as
\be
\mlr \to \Om_{\rm L} \mlr \Om_{\rm R}^{\dagger},
\;\;\;\;
\mrl \to \Om_{\rm R} \mrl \Om_{\rm L}^{\dagger}.
\ee
We will also encounter the field
strengths and covariant derivatives
\be
F_{\mu\nu}^{\rm L,R} = [D_{\mu}^{\rm L,R},D_{\nu}^{\rm L,R}],
\quad
\Da_{\mu} \mlr = \dmu\mlr + v_{\mu}^{\rm L}\mlr - \mlr
v_{\mu}^{\rm R},
\ee
and $\Da_{\mu}\mrl = (\Da_{\mu}\mlr)^{\dagger}$.

The fermion fields form an anomalous representation of the gauge
group and consequently
$W$ contains a chiral anomaly. It is not gauge invariant under the
full
$U(n)\times U(n)$ group but it will be so when restricted to the
gauge group of the Standard Model.

The effective action can be split into terms that are even and odd
under the pseudo-parity transformation $v^{\rm L}\leftrightarrow
v^{\rm R}$, $\mlr \leftrightarrow \mrl$,
\be
W = W^+ + W^-.
\ee
The `normal parity' component $W^+$ is formally identical to the
effective action of a vector-like model (see e.g.\ \cite{Salcedo:2000hp}).
We are interested in the
`abnormal parity' component $W^-$, since it is odd in the number
of $\gm_5$ matrices and will contain the leading CP-violating
terms. It contains the anomalous representation of the $U(n)\times
U(n)$ gauge group and can be written in the form
\be
W^- = \Gm_{\rm gWZW} + W^-_{\rm c},
\ee
where $\Gm_{\rm gWZW}$ is an extended gauged Wess-Zumino-Witten
(WZW) action that contains the chiral anomaly. The
remainder $W^-_{\rm c}$ is $U(n)\times U(n)$ gauge-invariant. When
we specialize the gauge fields to those of the Standard Model, for
which the fermion content is an anomaly-free representation of the
$U(1)\times SU(2) \times SU(3)$ gauge group 
\cite{Bouchiat:1972iq,Gross:1972pv,D'Hoker:1984ph}, 
$\Gm_{\rm gWZW}$ becomes
also gauge invariant (appendix \ref{appB}).

We start with $W^-_{\rm c}$. Using an
elegant and powerful notation \cite{Salcedo:2000hp,Salcedo:2000hx} Salcedo obtains
$W^-_{\rm c}$ in the condensed form
\be
W^-_{\rm c}[v,m] =
\frac{1}{48\pi^2}\intx \ep_{\kp\lm\mu\nu} {\rm tr}\left[
N_{123}\sDa_{\kp}\sm\, \sDa_{\lm}\sm\, \sF_{\mu\nu} + N_{1234}
\sDa_{\kp} \sm\, \sDa_{\lm}\sm\,
\sDa_{\mu}\sm\, \sDa_{\nu}\sm
\right].
\label{not1}
\ee
Here $N_{123}$, is a function of $\sm_1$, $\sm_2$ and
$\sm_3$, and similarly for
$N_{1234}$, in which the subscripts indicate the position where the
matrices $\sm$ are to be inserted in the trace;
$\sm$ and $\sF_{\mu\nu}$ are to be replaced by $\mlr$ or $\mrl$ and
$F_{\mu\nu}^{\rm R}$ or $F_{\mu\nu}^{\rm L}$,
with $\sDa$ the appropriate covariant derivative, such that a
gauge-invariant expression results. See \cite{Salcedo:2000hp,Salcedo:2000hx} for a full
exposition of the notation. To see how this works, consider the
first term in (\ref{not1}) with
$N_{123}$ replaced by the monomial $\sm_1\, \sm_2^2\, \sm_3^3$:
\bea
\tr \left[
\sm_1\, \sm_2^2\, \sm_3^3\, \sDa_{\kp}\sm\, \sDa_{\lm}\sm\, \sF_{\mu\nu}
\right]
&\equiv&
\tr\left[
\sm\, \sDa_{\kp}\sm\, \sm^2\, \sDa_{\lm}\sm\, \sm^3\, \sF_{\mu\nu}
\right]
\nonumber\\
&\equiv&
\half\tr\left[
\mrl \Da_{\kp}\mlr\,\mrl\mlr\,\Da_{\lm}\mrl\,
\mlr\mrl\mlr\,F_{\mu\nu}^{\rm R}
\right]
\nonumber\\&&
-({\rm L\leftrightarrow R}).
\label{not2}
\eea
More general functions $N_{123}$ are dealt with by going to a
basis in which $\mlr$ and $\mrl$ reduce to positive
diagonal matrices $d$. This can be achieved by making a `polar
decomposition' $\mlr = PU$ in which $U$ is unitary and $P$ is
hermitian and positive, and then diagonalize $P$, $\UL^{\dagger} P\UL =
d$, or $P = \UL d\, \UL^{\dagger}$, which leads to
\be
\mlr = \UL d\, \UR^{\dagger},
\quad
\mrl = \UR d\, \UL^{\dagger},
\ee
with $\UR^{\dagger} = \UL^{\dagger}U$. We also have
\be
\mlr\mrl = \UL d^2\, \UL^{\dagger},
\quad
\mrl\mlr = \UR d^2\, \UR^{\dagger},
\ee
etc. Even factors of $m$ have identical L or R labels left and
right. It follows that (\ref{not2}) can be written  in the form
\bea
&&\half{\rm tr}\left[
d\,\UL^{\dagger}\Da_{\kp}\mlr\UR\,d^2\,\UR^{\dagger}\Da_{\lm}\mrl\UL\,
d^3\,\UR^{\dagger}F_{\mu\nu}^{\rm R}\UR\right] -({\rm
L\leftrightarrow R})
\nonumber\\
&=&
\half\sum_{jkl}d_jd_k^2d_l^3 (\Da_{\kp}\mlr)_{jk}
(\Da_{\lm}\mrl)_{kl}(F_{\mu\nu}^{\rm R})_{lj} -({\rm
L\leftrightarrow R}),
\label{not3}
\eea
where $j,k,l = 1,\cdots,n$ are labels in the diagonal basis,
$(d)_{jk}=d_j\dl_{jk}$, and
\be
(F_{\mu\nu}^{\rm R})_{lj}=(\UR^{\dagger}F_{\mu\nu}^{\rm
R}\UR)_{lj},
\;\;
(\Da_{\kp}\mlr)_{jk} = (\UL^{\dagger}\Da_{\kp}\mlr\UR)_{jk},
\;\;
(\Da_{\lm}\mrl)_{kl} = (\UR^{\dagger}\Da_{\lm}\mrl\UL)_{kl}.
\ee
Since even factors of $m$ do not change L into R or R into L,
(\ref{not2}), \ldots, (\ref{not3}) can be generalized to
\be
\tr\left[
\sm_1^p\, \sm_2^q\, \sm_3^r\, \sDa_{\kp}\sm\, \sDa_{\lm}\sm\, \sF_{\mu\nu}
\right]
=
\half\sum_{jkl}d_j^p d_k^q d_l^r (\Da_{\kp}\mlr)_{jk}
(\Da_{\lm}\mrl)_{kl}(F_{\mu\nu}^{\rm R})_{lj}
 -({\rm L\leftrightarrow R}),
\ee
provided that the integers $p$ and $r$ are odd and $q$ is even.

The function $N_{123}$ is given in \cite{Salcedo:2000hx} and we have copied it
into appendix A. It is invariant under the simultaneous
sign flips $\sm_a \to - \sm_a$,
$a=1,2,3$. We decompose it into terms even and odd in
$\sm_1$, \dots, $\sm_3$:
\bea
N_{123} &=& f^{(0)} + f^{(12)} \sm_1\sm_2 + f^{(23)} \sm_2\sm_3 +
f^{(13)}\sm_1\sm_3
\\
&\equiv& N^{(0)} + N^{(12)} + N^{(23)} + N^{(13)},
\eea
where the $f$s are even functions,
e.g.\ $N^{(12)}(\sm_1,\sm_2,\sm_3) =
f^{(12)}(\sm_1^2,\sm_2^2,\sm_3^2)\sm_1\sm_2$. The first term in
the trace in (\ref{not1}) can then be written in the more explicit
form
\bea
{\rm tr}\left[
N_{123}\sDa_{\kp}\sm\, \sDa_{\lm}\sm\, \sF_{\mu\nu}\right]
&=&
\half\sum_{jkl}\left[
N^{(0)}_{jkl} (\Da_{\kp}\mrl)_{jk} (\Da_{\lm}\mlr)_{kl}
(F_{\mu\nu}^{\rm R})_{lj}
\right.
\nonumber\\&& \mbox{} +
N^{(12)}_{jkl} (\Da_{\kp}\mlr)_{jk}(\Da_{\lm}\mlr)_{kl}
(F_{\mu\nu}^{\rm R})_{lj}
\nonumber\\&& \mbox{} +
N^{(23)}_{jkl} (\Da_{\kp}\mrl)_{jk}
(\Da_{\lm}\mrl)_{kl}(F_{\mu\nu}^{\rm R})_{lj}
\nonumber\\&& \left. \mbox{} +
N^{(13)}_{jkl} (\Da_{\kp}\mlr)_{jk} (\Da_{\lm}\mrl)_{kl}
(F_{\mu\nu}^{\rm R})_{lj}
\right]
 -({\rm L\leftrightarrow R}),
\label{not4}
\eea
where
\be
N^{(12)}_{jkl} = N^{(12)}(d_j,d_k,d_l),
\label{Njkl}
\ee
etc. The second term involving $N_{1234}$ can be treated in
similar fashion but we shall postpone this for later.

\section{Application to the Standard Model}
\label{appSM}
We write the fermion part of the SM action, extended with
right-handed neutrino fields, in the form
\be
S_{\rm F}=\intx
\Psb\left\{\gm_{\mu}
\left[\dmu-iA_{\mu}\PL-iG_{\mu}-i(\YL\PL + \YR\PR)B_{\mu}
\right]
+\Ph\Lm\PR + \Lm^{\dagger}\Phd\PL\right\}\Ps.
\ee
Here $\Ps$ is a $4{\rm (Dirac)}\times n$-component spinor, where
$n=2{\rm(isospin)}\times (3{\rm (color)}+1) \times 3{\rm(family)} = 24$
(the leptons are represented by the `1').
The gauge fields are taken to be hermitian: $B_{\mu}$ for $U(1)$,
$A_{\mu}$ for $SU(2)$ and $G_{\mu}$ for $SU(3)$. The matrix fields
$A_{\mu}$ and $G_{\mu}$ and also the coupling matrices $Y$ and $\Lm$
are embedded into the grand structure in the usual
tensor product fashion: the spinor field has components
(suppressing the Dirac indices)
\be
\Ps^k,
\quad
k=(i,c,f),
\quad
i\in \{u,d\},
\quad
c\in \{1,2,3\},
\quad
f\in \{1,2,3\},
\label{index}
\ee
with (weak) isospin index $i$, color index $c$ and family index $f$,
for quarks, and of course no color index for the leptons.
The $SU(2)$ gauge fields can be written in terms of
Pauli matrices as
$A_{\mu} = A_{\mu}^a \ta_a/2$, with
$(\ta_a)_{kk'} = (\ta_a)_{ii'}\dl_{cc'}\dl_{ff'}$. Similarly
the $SU(3)$ fields are embedded as
$(G_{\mu})_{kk'} = (G_{\mu})_{cc'} \dl_{ii'}\dl_{ff'}$.
We find it convenient to make the $SU(2)$ structure explicit:
\be
A_{\mu} = \half\, \ta_a\,  A_{\mu}^a,
\quad
\ta_1 = \left(\begin{array}{cc}0&1\\1&0\end{array}\right),
\;\;
\ta_2 = \left(\begin{array}{cc}0&-i\\i&0\end{array}\right),
\;\;
\ta_3 = \left(\begin{array}{cc}1&0\\0&-1\end{array}\right).
\ee
The $Y$ are diagonal matrices representing the $U(1)$ hypercharges:
\be
\YL =  \left(\begin{array}{cc}\frac{1}{6}&0\\0&\frac{1}{6}\end{array}\right)
\pi_q +
\left(\begin{array}{cc}-\half&0\\0&-\half\end{array}\right)
\pi_{\ell},
\quad
\YR =  \left(\begin{array}{cc}\frac{2}{3}&0\\0&-\frac{1}{3}\end{array}\right)
\pi_q +
\left(\begin{array}{cc}0&0\\0&-1\end{array}\right)
\pi_{\ell},
\ee
where $\pi_q$ and $\pi_{\ell}$ project respectively onto the quark and
lepton labels.
The Higgs field is in matrix form. In terms of the $SU(2)$ Higgs-doublet
$(\vr^u,\vr^d)^T$ it reads
\be
\Ph = \left(\begin{array}{cc}\vr^{d\ast}&\vr^u\\
-\vr^{u\ast}&\vr^d\end{array}\right),
\quad
\Ph \to \Om \Ph e^{-i\om \ta_3/2}
,
\ee
where we have also indicated its behavior under gauge transformations,
$e^{i\om/6}\in U(1)$, $\Om\in SU(2)$. Note that, since $\ta_3/2=\YR-\YL$
and $\YL$ commutes with $\Ph$,
this can also be written as $\Ph\to\Om_{\rm L}\Ph\Om_{\rm R}^{\dagger}$,
with
$\Om_{\rm L} = e^{i\om \YL}\Om$, $\Om_{\rm R} = e^{i\om\YR}$.
The matrix $\Lm$ represents the Yukawa couplings.
Its $SU(2)$ structure is given by
\be
\Lm =
\left(\begin{array}{cc}\Lm^u_q&0\\0&\Lm^d_q\end{array}\right)
\pi_q +
\left(\begin{array}{cc}\Lm^u_\ell&0\\0&\Lm^d_\ell\end{array}\right)
\pi_{\ell},
\label{Lambda1}
\ee
where the $\Lm^{u}_{q}, \ldots, \Lm^d_\ell$ are non-trivial matrices
in family space.
Note that we have not included a Majorana mass term for the right-handed
neutrino fields (often invoked for the see-saw mechanism), since this does
not fit straight-away into the $\psb\cdots\ps$ form assumed in Salcedo's
calculation of the effective action.

It follows that the fields in the previous section are realized as
\be
v_{\mu}^{\rm L} = -i\YL B_{\mu} -iA_{\mu} -iG_{\mu},
\quad
v_{\mu}^{\rm R} = -i \YR B_{\mu} -i G_{\mu},
\quad
\mlr = \Ph\Lm,
\quad
\mrl = \Lm^{\dagger}\Phd,
\label{SMfields}
\ee
and
\be
F^{\rm L}_{\mu\nu} = -iA_{\mu\nu} -i \YL B_{\mu\nu} -i G_{\mu\nu},
\quad
F^{\rm R}_{\mu\nu} = -i\YR B_{\mu\nu} -i G_{\mu\nu},
\ee
\be
\Da_{\mu}\mlr = \left(\dmu\Ph - iA_{\mu}\Ph + i \Ph \half \ta_3 B_{\mu}\right)
\Lm,
\quad
\Da_{\mu}\mrl = \left( \Da_{\mu}\mlr\right)^{\dagger},
\ee
with $A_{\mu\nu} = \dmu A_{\nu} - \dnu A_{\mu} -i[A_{\mu},A_{\nu}]$, etc.
The diagonal basis used in (\ref{not4}) is obtained by diagonalizing $\Lm$
and futhermore transforming to the unitary gauge in which
$\Ph = h\, \unity$ ($\sqrt{2} h$ is the standard-normalized Higgs field):
\bea
\Lm &=& \VL\lm \VR^{\dagger},
\quad
\lm = \left(\begin{array}{cc}\lm^u_q&0\\0&\lm^d_q\end{array}\right)
\pi_q +
\left(\begin{array}{cc}\lm^u_\ell&0\\0&\lm^d_\ell\end{array}\right)
\pi_{\ell},
\\
\VL &=& \left(\begin{array}{cc}\VL^u&0\\0&\VL^d\end{array}\right)_q
\pi_q +
\left(\begin{array}{cc}\VL^u&0\\0&\VL^d\end{array}\right)_{\ell}
\pi_{\ell},
\eea
and similar for $\VR$; furthermore
\be
\Ph = \Om h,\;\;\;\; \Om\in SU(2),
\ee
and so
\be
\mlr = \UL\, d\, \UR^{\dagger},
\;\;\;\; \UL = \Om \VL,
\;\;\;\;\UR = \VR,
\quad
d = h\, \lm.
\label{ev}
\ee
The $\lm$ are non-trivial diagonal matrices in family space,
$\lm^u_q = {\rm diag}(\lm_u,\lm_c,\lm_t)$,
$\lm^d_q = {\rm diag}(\lm_d,\lm_s,\lm_b)$, and similar for the leptons.
In the following we will concentrate on the quark contribution to the
effective action (the lepton contribution is analogous),
and omit the subscripts $q$ and $\ell$ if there is no danger of confusion.

The matrix elements of the covariant derivatives
entering in (\ref{not4}) in the diagonal basis are now given by
\bea
\UL^{\dagger} \Da_{\mu}\mlr \UR &=&
\VL^{\dagger} (h^{-1}\dmu h -i W_{\mu} + i B_{\mu} \ta_3/2)\VL d
\\
&\equiv& -i C_{\mu} d,
\\
\UR^{\dagger} \Da_{\mu}\mrl \UL &=&
i d\, C_{\mu}^{\dagger} ,
\\
C_{\mu} &=& ih^{-1}\dmu h + W^a_{\mu} \tilde \ta_a/2
- B_{\mu} \ta_3/2,
\label{Cdef}
\eea
where $W_{\mu}$ is the $SU(2)$ gauge field in unitary gauge,
\be
W_{\mu} = \Om^{\dagger}A_{\mu}\Om + i\Om^{\dagger}\dmu\Om,
\ee
and
\be
\tilde\ta_1 =
\left(\begin{array}{cc}0&V\\V^{\dagger}&0\end{array}\right),
\quad
\tilde\ta_2 =
\left(\begin{array}{cc}0&-iV\\iV^{\dagger}&0\end{array}\right),
\quad
\tilde \ta_3 = \ta_3,
\quad
V = \VL^{u\dagger}\VL^d.
\ee
Here $V$ is the celebrated CKM matrix.
Similarly,
\bea
\UL^{\dagger} F^{\rm L}_{\mu\nu}\UL
&=&
-i \tilde W_{\mu\nu} -i\YL B_{\mu\nu} -i G_{\mu\nu},
\\
\UR^{\dagger} F^{\rm R}_{\mu\nu} \UR
&=&
-i \YR B_{\mu\nu} -i G_{\mu\nu} =  F^{\rm R}_{\mu\nu},
\eea
with
\be
\tilde W_{\mu\nu} = W_{\mu\nu}^a \tilde \ta_a/2.
\ee
Note that in (\ref{Cdef}) the combination $W^3_{\mu}-B_{\mu}=Z_{\mu}$,
the $Z$ field with coupling constants absorbed,
\be
B_\mu= A_\mu - \sin^2\theta_{\rm W} Z_\mu,
\quad
W^3_\mu = A_\mu+\cos^2\theta_{\rm W} Z_\mu,
\ee
with $A_{\mu}$ is the photon field (with electro-magnetic
coupling $e$ absorbed) and
$\theta_{\rm W}$ the Weinberg angle.

Consider now the $N^{(0)}$ contribution in (\ref{not4}). With
the above specialization to the SM this becomes, including
also the epsilon tensor from (\ref{not1}),
\bea
&&
\ep_{\kp\lm\mu\nu}\half \sum_{jkl} \left[
N^{(0)}_{jkl}\, d_j (C^{\dagger}_{\kp})_{jk}
(C_{\lm})_{kl}\, d_l\, (-i\YR B_{\mu\nu} -iG_{\mu\nu})_{lj}
\right. \nonumber\\ && \left.\mbox{}
- N^{(0)}_{jkl} (C_{\kp})_{jk}\, d_k^2\,
(C^{\dagger}_{\lm})_{kl}\,
(-i \tilde W_{\mu\nu} - i\YL B_{\mu\nu} -i G_{\mu\nu})_{lj}
\right].
\label{not11}
\eea
Firstly, we observe that the $SU(3)$ fields do not contribute, since
the egenvalues $d_j$ and hence also
$N^{(0)}_{jkl}$ are color-independent, $(C_{\mu})_{kk'} \propto
\dl_{cc'}$ (cf.\ (\ref{index})),
and $G_{\mu\nu}$ is traceless.
Secondly, because of the property (cf.\ (\ref{propN3}))
\be
N^{(0)}_{jkl} = - N^{(0)}_{lkj},
\label{n0symm2}
\ee
we have $N^{(0)}_{jkj} = 0$, and consequently
also the fields $B_{\mu\nu}$ and
$W^3_{\mu\nu}$ involving the diagonal generators
$Y_{\rm L,R}$ and $\ta_3$ drop out.
So we are left with the off-diagonal contribution from $\tilde W_{\mu\nu}$:
\be
\ep_{\kp\lm\mu\nu}\frac{i}{4} \sum_{jkl}
 N^{(0)}_{jkl} d_k^2\, (C_{\kp})_{jk}\,
(C^{\dagger}_{\lm})_{kl}\,
(\tilde\ta_1 W^1_{\mu\nu} + \tilde\ta_2 W^2_{\mu\nu})_{lj}.
\label{1}
\ee
Using interchange of dummy indices $\kp\leftrightarrow\lm$ this can also
be written as
\be
\ep_{\kp\lm\mu\nu}\frac{-i}{4} \sum_{jkl}
 N^{(0)}_{jkl} d_k^2\, (C_{\lm})_{jk}\,
(C^{\dagger}_{\kp})_{kl}\,
(\tilde\ta_1 W^1_{\mu\nu} + \tilde\ta_2 W^2_{\mu\nu})_{lj},
\ee
and using furthermore
$j\leftrightarrow l$, the property (\ref{n0symm2}) and the hermiticity
of the $\tilde\ta_a$, this can be rewritten
as
\be
\ep_{\kp\lm\mu\nu}\frac{i}{4} \sum_{jkl}
 N^{(0)}_{jkl} d_k^2\, (C_{\kp}^*)_{jk}\,
(C^{\dagger\ast}_{\lm})_{kl}\,
(\tilde\ta_1^* W^1_{\mu\nu} + \tilde\ta_2^* W^2_{\mu\nu})_{lj}.
\label{2}
\ee
Combining (\ref{1}) and (\ref{2}), it follows that the expression is purely
imaginary,
\be
\ep_{\kp\lm\mu\nu}\frac{i}{4} \sum_{jkl}
 N^{(0)}_{jkl} d_k^2\, {\rm Re}\left[(C_{\kp})_{jk}\,
(C^{\dagger}_{\lm})_{kl}\,
(\tilde\ta_1 W^1_{\mu\nu} + \tilde\ta_2 W^2_{\mu\nu})_{lj}\right],
\ee
which is a general property of the pseudoparity-odd contribution to the
euclidean effective action.
We can now examine the type of contributions:
\begin{itemize}
\item[$(i)$]
$C_{\kp}\to W^a_{\kp}\tilde\ta_a/2$, $C_{\lm}\to W^b_{\lm}\tilde\ta_b/2$
leads to
\be
i
\ep_{\kp\lm\mu\nu}
W^a_{\kp}\, W^b_{\lm}\, W^c_{\mu\nu}
\, n_{abc}^{(0)},
\ee
with
\be
n^{(0)}_{abc} =
\frac{1}{16}
\sum_{jkl}
 N^{(0)}_{jkl} d_k^2\, {\rm Re}\left[(\tilde\ta_a)_{jk}\,
(\tilde\tau_b)_{kl}\,
(\tilde\ta_c)_{lj}\right].
\ee
We need to investigate $n^{(0)}_{abc}$.
As seen above, it is nonzero only for $c=1,2$.
Suppose $a=3$. Then $b$ has to be 1 or 2 because of the off-diagonality of
$\tilde\ta_c$.
Similarly, if $b=3$ then only $a=1,2$ can give a non-zero contribution.
The two cases $a=3$ or $b=3$ lead essentially to the same result
and we continue with $a=3$. In this case
only $j=k$ contributes because $\ta_3$ is diagonal.
In the notation (\ref{index}),
let $k=(i,c,f)$, $l=(i',c',g)$.
The matrices $\tilde\ta_a$ are color-diagonal,
so we only need $c=c'$. The eigenvalues $d_j$ do not depend on color,
$d_k = d_{if}$, and the summation over $c$ just gives a factor 3.
For $b=c=1$ this gives
\be
n^{(0)}_{311} = \frac{3}{16} \sum_{fg} \left[ N^{(0)}_{uf,uf,dg} d_{uf}^2
{\rm Re}\left(V_{fg}V^{\dagger}_{gf}\right)
- N^{(0)}_{df,df,ug} d_{df}^2 {\rm Re}\left(V^{\dagger}_{fg}V_{gf}\right)
\right],
\label{n0311}
\ee
and the same for $b=c=2$, leading to a contribution
\be
i
\ep_{\kp\lm\mu\nu}
W^3_{\kp} \left(W^1_{\lm}\, W^1_{\mu\nu} + W^2_{\lm}\, W^2_{\mu\nu}\right)
n_{311}^{(0)}.
\label{cpOK1}
\ee

For $b=1$ and $c=2$ we get
\bea
n^{(0)}_{312} &=& -\frac{3}{16}\sum_{fg} \left[N^{(0)}_{uf,uf,dg} d_{uf}^2
{\rm Im}\left(V_{fg}V^{\dagger}_{gf}\right)
+ N^{(0)}_{df,df,ug} d_{df}^2
{\rm Im}\left(V^{\dagger}_{fg}V_{gf}\right)
\right]
\nonumber\\
&=& 0,
\eea
since the imaginary part is zero,
and `minus zero' for $b=2$, $c=1$, i.e.\ the coefficient of
\be
i
\ep_{\kp\lm\mu\nu}
W^3_{\kp} (W^1_{\lm}\, W^2_{\mu\nu} - W^2_{\lm}\, W^1_{\mu\nu})
\label{cpnOK}
\ee
vanishes.

\item[$(ii)$]
$C_{\kp}\to i h^{-1}\partial_{\kp} h$, $C_{\lm}\to W^b_{\lm}\tilde\ta_b/2$
leads to
\be
i\ep_{\kp\lm\mu\nu}
h^{-1}\partial_{\kp}h\, W^b_{\lm} W^c_{\mu\nu}\,
n^{(0)}_{bc},
\label{cpOK2}
\ee
with
\be
n^{(0)}_{bc} = - \frac{1}{8}
\sum_{jkl}
 N^{(0)}_{jkl} d_k^2\, \dl_{jk}\, {\rm Im}\left[
(\tilde\ta_b)_{kl}\,
(\tilde\ta_c)_{lj}\right].
\label{n0bc}
\ee
In this case we find non-zero results only for $b=1$, $c=2$ and $b=2$, $c=1$.
\end{itemize}

The examples above show the general feature that also holds for the other
contributions involving $N^{(12)}$, $N^{(23)}$ and $N^{(13)}$ in (\ref{not4}):
CP conserving terms such as (\ref{cpOK1}), and (\ref{cpOK2}) with $b\neq c = 1,2$,
survive, but all the CP-violating contributions like (\ref{cpnOK})
vanish.\footnote{Under CP the fields transform as
$\Ph(x)\to\Ph^*(Px)$, $A_{\mu}(x)\to-P_{\mu\nu} A_{\mu}^T(Px)$,
$B_{\mu}(x)\to -P_{\mu\nu} B_{\nu}(Px)$, with
$P = {\rm diag}(1,-1,-1,-1)$. Specifically in unitary gauge,
$h(x)\to h(Px)$, $W^{1,3}_{\mu} \to - P_{\mu\nu} W^{1,3}_{\nu}(Px)$,
$W^2_{\mu}(x)\to +P_{\mu\nu} W^2_{\nu}(Px)$,
$Z_{\mu}(x)\to - P_{\mu\nu}Z_{\nu}(Px)$ and
$A_{\mu}(x)\to - P_{\mu\nu}A_{\nu}(Px)$. }
The reason is evidently that there are
not enough CKM matrices present in the above expressions to be
able to construct even the minimal CP-violating invariant under
phase redefinitions, (\ref{J}), which is of fourth order in
$V,V^*$.

It is now also not difficult to see that the $N_{1234}$
contribution in (\ref{not1}) cannot contain CP-violating terms in
the Standard Model case. For example, the $N^{(0)}_{jklm}$
contribution leads to terms of the form
\be
\ep_{\kp\lm\mu\nu}\sum_{jklm}N^{(0)}_{jklm} d_j^2\, d_l^2\,(C^{\dagger}_{\kp})_{jk}
(C_{\lm})_{kl}(C^{\dagger}_{\mu})_{lm}(C_{\nu})_{mj}
\ee
Using $N^{(0)}_{jklm} = -N^{(0)}_{lkjm}$, which follows from the
properties (\ref{propN4}) and the fact that $N^{(0)}_{jklm}$ is an
even function of $d_j$, \ldots, $d_m$, this expression can be
shown to be purely imaginary. Choosing from the $C$ the purely gauge-field
contribution leads to
\be
i\ep_{\kp\lm\mu\nu} W^a_{\kp}W^b_{\lm}W^c_{\mu}W^d_{\nu}
\frac{1}{16}\sum_{jklm}N^{(0)}_{jklm}d_j^2\,d_l^2\,{\rm Im}
\left[(\tilde\ta_a)_{jk}(\tilde\ta_b)_{kl}(\tilde\ta_c)_{lm}(\tilde\ta_d)_{mj}\right],
\ee
where the $W^3$ field can be replaced by the $Z$ field. The above
expression appears to contain enough factors of $V$ and
$V^{\dagger}$ to be able to make up the invariant $J$. However,
the $\ep$-tensor projects this contribution to zero as there are
not enough independent four-vectors. Next, assume the Higgs field
contribution in one of the $C$, say
$C_{\nu}\to ih^{-1}\dnu h$, which leads to
\be
i\ep_{\kp\lm\mu\nu} W^a_{\kp}W^b_{\lm}W^c_{\mu}\,h^{-1}\dnu h\,
\frac{1}{8}\sum_{jklj}N^{(0)}_{jklm}d_j^2\,d_l^2\,{\rm Re}
\left[(\tilde\ta_a)_{jk}(\tilde\ta_b)_{kl}(\tilde\ta_c)_{lm}\right],
\ee
which violates CP. The $\ep$-tensor requires $a,b,c$ to be a
permutation of 1,2,3. Since $\tilde\ta_3$ does not contain
$V$ and $\ta_2$ is imaginary this implies taking the imaginary part of a
phase-invariant combination of two only $V$'s, which is zero.

We conclude that to fourth order in the derivative expansion,
there are no CP-violating terms in $W^-_{\rm c}$.
A similar analysis and conclusion applies to the analog mixing matrix
in the lepton contribution to the $W^-_{\rm c}$,
provided there is no Majorana neutrino mass term.

\section{Magnitude of CP violation}
\label{magn}
To find CP violation coming from the CKM matrix we need to go to
higher order in the derivative expansion. For example, we
anticipate in $W^-_{\rm c}$ a sixth-order term of the form
\be
\ep_{\kp\lm\,u\nu} \tr\left[
N'_{1234} \sF_{\kp\lm} \sF_{\mu\nu} \sDa_{\rh}\sm\, \sDa_{\rh}\sm\right].
\ee
Decomposing as before $N'_{1234} = N^{\prime(0)}_{1234} + \cdots$, in which
$N^{\prime(0)}_{1234}$ depends only on $\sm^2$, this contains the
contribution
\be
\frac{-i}{2}\sum_{jklm} N^{\prime(0)}_{jklm} {\rm Im}\left[
(F^{\rm L}_{\kp\lm})_{jk} (F^{\rm L}_{\mu\nu})_{kl} (\Da_{\rh}
\mlr)_{lm} (\Da_{\rh} \mrl)_{mj}\right] - ({\rm L}\leftrightarrow
{\rm R}),
\ee
where we also used the fact that $W^-$ is imaginary (assuming the
$N$-functions to be real as for the fourth-order terms).
The purely $SU(2)$-field contribution is then given by
\be
\frac{-i}{32}\ep_{\kp\lm\mu\nu} W^a_{\kp\lm} W^b_{\mu\nu} W^c_{\rh} W^d_{\rh}
\sum_{jklm} N^{\prime (0)}_{jklm}\, d_m^2\,
{\rm Im}\left[
(\tilde\ta_a)_{jk} (\tilde\ta_b)_{kl} (\tilde\ta_c)_{lm} (\tilde\ta_d)_{mj}
\right].
\ee
There are several CP-violating contributions, e.g.\ the ones with
$a=b=1,2$, $c=d=1,2$, are proportional to
\be
\sum_{f,g,h,i}\left\{ N^{\prime(0)}_{uf,dg,uh,di}\, d_{di}^2\,
{\rm Im}\left[V_{fg}V^*_{hg}V_{hi}V^*_{fi} \right]
+ N^{\prime(0)}_{df,ug,dh,ui}\, d_{ui}^2\,
{\rm Im}\left[V^*_{gf}V_{gh}V^*_{ih}V_{if}\right]\right\},
\label{cpviol6}
\ee
in which the expected rephasing invariant $J$ appears. Generically
we do not expect these contributions to vanish. Their explicit
calculation appears a very complicated task. However, we now argue
that they are {\em not} accompanied by the tiny product of Yukawa
couplings in (\ref{dlcpest2}).

A striking feature of the $N_{123}$ and $N_{1234}$ of the
fourth-order contribution is the fact that they are homogeneous
functions (cf.\ appendix
\ref{appA}):
\be
N(s\sm_1,s\sm_2,s\sm_3) = s^{-2} N(\sm_1,\sm_2,\sm_3),
\quad
N(s\sm_1,s\sm_2,s\sm_3,s\sm_4) = s^{-4}
N(\sm_1,\sm_2,\sm_3,\sm_4).
\ee
Recalling (cf.\ \ref{ev}) that the eigenvalues of $\sm$ are given
by $d_j = h\lm_j$, it follows that in expressions such as
(\ref{not11}) or (\ref{n0311}) (recall also that
$N^{(0)}_{jkl} = N^{(0)}(d_j,d_k,d_l)$, as in (\ref{Njkl})),
the Higgs field $h$ hidden in
$d$ drops out altogether. It only occurs via its derivative
in the combination $h^{-1}\dmu h$ as in (\ref{cpOK2}).
Furthermore, the overall scale of the Yukawa couplings does not
matter: rescaling
$\lm_j \to s \lm_j$ does not change these expressions. This can be
seen clearly from the explicit expression for the combination
$N^{(0)}_{jkl} d_k^2$ appearing e.g.\ in (\ref{not11}), (\ref{n0311})
and (\ref{n0bc}): in terms of $d_j\equiv x$, $d_k\equiv y$,
$d_l\equiv z$ it reads (cf.\ \ref{defN0}))
\bea
N^{(0)}_{jkl} d_k^2 &=&2\left\{
\frac{2x^4+2z^4 -2x^2 z^2- x^2 y^2 -z^2 y^2}{
(x^2-y^2)(z^2-y^2)(x^2-z^2)}
\right.\\&&\mbox{}
+\left[\frac{x^4y^2+z^4y^2+x^2
z^4-3x^4z^2}{(x^2-y^2)^2(x^2-z^2)^2}
\log\frac{x^2}{y^2}
\right.\\&&\mbox{}\left.\left.
-(x\leftrightarrow z)\frac{\mbox{}^{\mbox{}}}{\mbox{}^{\mbox{}}}\right]
\right\}y^2.
\eea
Hence, the fourth-order contribution to
$W^-$ is invariant under
$\lm_j \to s\lm_j$. Such an
insensitivity to the overall scale of the $\lm$'s may very well be
present also in the CP-violating terms in higher orders of the
derivative expansion, such as anticipated in (\ref{cpviol6}). 
This strongly suggests that the product of
$\lm$'s should be ignored in rough estimates of the magnitude of
CP violation. 
For example,
\be
\frac{(\lm_{u}^{2}-\lm_{c}^{2}) (\lm_{c}^{2}-\lm_{t}^{2})
(\lm_{t}^{2}-\lm_{u}^{2}) (\lm_{d}^{2}-\lm_{s}^{2})
(\lm_{s}^{2}-\lm_{b}^{2}) (\lm_{b}^{2}-\lm_{d}^{2})}
{(\lm_{u}^{2}+\lm_{c}^{2}) (\lm_{c}^{2}+\lm_{t}^{2})
(\lm_{t}^{2}+\lm_{u}^{2}) (\lm_{d}^{2}+\lm_{s}^{2})
(\lm_{s}^{2}+\lm_{b}^{2}) (\lm_{b}^{2}+\lm_{d}^{2})}
\simeq 0.99.
\ee

The reasoning above does not apply to the case of finite
temperature $T$, for which $T$ provides a new scale. Salcedo's
results used here hold only for zero temperature. For example, at
sufficiently high temperature we may expect the appearance of
hard-thermal-loop masses $m_{\rm th}$, via
$\lm_j h \to \lm_j h + m_{\rm th}^j$. For quarks the QCD
contribution dominates, $m_{\rm th}^j \approx g_{\rm s} T/\sqrt{6}$, with
$g_{\rm s}$ the strong ($SU(3)$) gauge coupling (see e.g.\ \cite{LeBellac96}).
There is no reason to expect the thermal masses to cancel
completely in the denominators and the finite-temperature estimate
(\ref{dlcpest1}) may still hold truth.

\section{The extended gauged Wess-Zumino-Witten action}
\label{gWZW}

We now turn to the $\Gm_{\rm gWZW}$ part of the effective action.
It is given in \cite{Salcedo:2000hx} using the
notation of differential forms, in addition to the notational conventions
already used in section \ref{salc}. The following one-forms
are introduced \cite{Salcedo:2000hx}:
\be
\sR = \sm^{-1}d\sm,
\quad \sL = \sm \, d \sm^{-1}.
\ee
In terms of these, Salcedo's extended gauged WZW action is given
by
\bea
\Gamma_{\rm gWZW}&=&
\frac{1}{48\pi^2}\int\tr\left( -\frac{1}{5}\sR^5\right)
\nonumber\\&&\mbox{}
+ \frac{1}{48\pi^2}\int\tr\left[
-(\sR^3+\sL^3)\sv
+\frac{1}{2}(\sR\sv)^2 +\frac{1}{2}(\sL\sv)^2
+\sR^2\sv\sm^{-1}\sv\sm +\sL^2\sv\sm\sv\sm^{-1}
\right. \nonumber \\ && \left.
+\sR\sm^{-1}\sv\sm d\sv +\sL\sm\sv\sm^{-1}d\sv
+(\sR+\sL)\sv^3
+\sR\sv\sm^{-1}\sv\sm\sv + \sL\sv\sm\sv\sm^{-1}\sv
\right. \nonumber \\ && \left.
+(\sR+\sL + \sm^{-1}\sv\sm + \sm\sv\sm^{-1}) \{\sv,d\sv\}
+\sm\sv\sm^{-1}\sv^3 +\sm^{-1}\sv\sm\sv^3
\right. \nonumber \\ && \left.
+\frac{1}{2}(\sm\sv\sm^{-1}\sv)^2
 \right].
\label{gWZW2}
\eea
The first integral is over a five-dimensional manifold which has
four-dimensional euclidean space-time as a boundary. The
second integral is over four-dimensional space-time. An alternative version
\cite{Salcedo:2000hx} that exhibits the properties under gauge transformations
more clearly is recalled in appendix \ref{appB}.

Because the SM reduction is gauge invariant we may use again the
unitary gauge, which makes it easier to deal with the factors of
$\sm^{-1}$. Consider for example
\bea
\int \tr\left[R\sm^{-1}\sv\sm d\sv\right]
&=&
\intx \ep_{\kp\lm\mu\nu}\,\tr\left[\sm^{-1}\partial_\kp\sm\,\sm^{-1}\sv_\lm
\sm\,\dmu\sv_\nu\right]
\label{ex0}\\
&\equiv& \intx \ep_{\kp\lm\mu\nu}\,\half\,
\tr\left[\mlr^{-1}\partial_\kp\mlr\,
\mlr^{-1}v_\lm^{\rm L}\mlr\, \dmu v_\nu^{\rm R}\right]
- ({\rm L} \leftrightarrow {\rm R}).
\nonumber
\eea
In the unitary gauge $\mlr = h\Lm$, $\mlr^{-1} = h^{-1} \Lm^{-1}$, and
$\mlr^{-1}\partial_\kp\mlr = \mrl^{-1}\partial_\kp\mrl = h^{-1}\partial_\kp h$.
This gives (cf.\ (\ref{SMfields}))
\bea
&&
(-i)^2\intx \ep_{\kp\lm\mu\nu}\,
\left\{
\half\,h^{-1} \partial_{\kp} h\, \tr\left[
\Lm^{-1}(G_\lm + W_\lm + \YL B_\lm)\Lm\,
\dmu(G_\nu +\YR B_\nu)\right]\right.
\nonumber\\&&\left.\mbox{}
-\half\, h^{-1} \partial_{\kp} h\, \tr\left[
\Lm^{\dagger -1}(G_\lm+\YR B_\lm)\Lm^{\dagger} \,
\dmu(G_\nu +W_\nu +\YL B_\nu)\right]\right\}
\nonumber\\
&& =0,
\label{ex}
\eea
where we used the fact that $\YR$ and $\YL$ commute with $\Lm$,
and partial integration. Evaluating all the terms this way we
find\footnote{Since there are an even number of $SU(2)$ doublets,
the first term in (\ref{gWZW2}) is an unobservable multiple of
$2\pi i$; it is evidently zero in the unitary gauge because of
antisymmetry in the differential form.}
\be
\Gm_{\rm gWZW} = 0.
\ee

An unsatisfactory aspect of this result is that, since total
derivatives have been dropped, the QCD $\theta$-term has been lost
as well. It is supposed to be produced by the chiral anomaly, upon
diagonalization of the quark mass-matrix $\propto \Lm_q$. To
recover $\theta$ terms, we initially allow $\Lm$ to be space-time
dependent in the reduction to the Standard Model. Then the purely
$SU(3)$ gauge-field contribution to (\ref{ex0}) produces factors
(cf.\ (\ref{Lambda1}))
\be
\tr_{\rm if}(\Lm^{-1}\partial_\kp\Lm\,\pi_q) =
\tr_{\rm f}({\Lm^u_q}^{-1}\partial_\kp\Lm^u_q) + (u\to d)
= \partial_\kp \tr_{\rm f}(\ln \Lm^u_q) + (u\to d)
= i\partial_\kp (\theta^u_q + \theta^d_q)
\ee
where $\theta^{u,d}_q = \arg\det  \Lm^{u,d}_q$ and $\tr_{\rm if}$ and
$\tr_{\rm f}$ are traces in isospin-family and in family space, respectively.
We assume that the
$\theta\to 0$ as $|x|\to\infty$ fast enough, initially, to allow for
partial integration without surface terms.
After removing $\partial_\kp$
from the $\theta$'s by partial integration they are taken to be constant.
We then recover the QCD $\theta$-term with $\theta = \theta^u_q + \theta^d_q$
from the terms linear in $L$ and $R$ (see also below).
In the complete case with also the $U(1)$ and $SU(2)$ gauge fields
present there may be also be contributions from the terms
non-linear in $L$ and $R$. To avoid such contributions we promote
only the phase of the total determinants of $\Lm_q$ and $\Lm_\ell$
to axion-like fields, writing
\be
\Lm^{-1}_q\partial_\kp\Lm_q
= i\partial_\kp\theta_q\frac{1}{n_{\rm if}}\,\unity,
\quad
\theta_q = \arg\det\Lm_q,
\label{chiintro}
\ee
and similar for $q\to \ell$;
here $n_{\rm if}=6$ is the number of families times the dimension of
isospin space.
With only two independent vectors,
$\partial_\kp\theta_q$ and $\partial_\kp\theta_\ell$,
the $\theta$ can appear only linearly in $\Gm_{\rm gWZW}$ because of
$\ep_{\kp\lm\mu\nu}$,
since quark and lepton contributions are not mixed.
We can implement (\ref{chiintro}) as $\Lm_q =\Lm'_q\, e^{i\theta_q/n_{\rm if}}$,
$\det \Lm'_q = 1$, with $\Lm'_q$ independent of $x$.
Effectively this implies
\be
\mlr^{-1}\partial_\kp\mlr \to  h^{-1}\partial_{\kp} h
+ i\partial_\kp \theta/n_{\rm if},
\quad
\mrl^{-1}\partial_\kp \mrl \to h^{-1}\partial_{\kp} h
- i\partial_\kp \theta/n_{\rm if},
\ee
with
\be
\theta =
\theta_q\,\pi_q + \theta_\ell\,\pi_\ell.
\ee
In addition to (\ref{ex}) we now also get the non-zero contribution
\bea
&&
- i\frac{1}{n_{\rm if}} \intx \ep_{\kp\lm\mu\nu}\,
\left\{
\half \tr\left[\partial_\kp\theta\,
(G_\lm + W_\lm + \YL B_\lm)
\dmu(G_\nu + \YR B_\nu)\right]\right.
\nonumber\\&&\left.\mbox{}
+\half\tr\left[\partial_\kp\theta\,
(G_\lm+\YR B_\lm) \,
\dmu(G_\nu +W_\nu +\YL B_\nu)\right]\right\}.
\eea
Collecting all the terms and making a partial integration to take
away the derivative from $\theta$ we get
\bea
\Gm_{\rm gWZW}&=&
i\frac{1}{48\pi^2}\intx
\ep_{\kp\lm\mu\nu}\,
\partial_\kp \frac{1}{n_{\rm if}}\, \tr \left\{\theta
[6 G_\lm\dmu G_\nu -i 4 G_\lm G_\mu G_\nu
\right.\nonumber\\&&\mbox{}
+ (W_\lm + \YL B_\lm)\dmu B_\nu + \YR B_\lm\dmu(W_\nu + \YL B_\nu)
\nonumber\\&&\mbox{}
-i (W_\lm+\YL B_\lm)(W_\mu + \YL B_\mu)(W_\nu + \YL B_\nu)
\nonumber\\&&\mbox{}
-i (W_\lm + \YL B_\lm) \YR B_\mu(W_\nu + \YL B_\nu)
\nonumber\\&&\mbox{}
\left.\left.
+ 2\YR^2 B_\lm\dmu B_\nu + 2(W_\lm + \YL B_\lm)\dmu(W_\nu + \YL B_\nu)
\right]\right\}.
\eea
Except for the first line, the order of the terms corresponds roughly to the
order of the terms in (\ref{gWZW2}).

We now let $\theta_{q,\ell}$ become constants. Then the first line takes the form
of the QCD $\theta$-term, since
\bea
&&\ep_{\kp\lm\mu\nu} \frac{1}{48\pi^2 n_{\rm if}}\, \partial_\kp
\tr \left\{\theta
[6 G_\lm\dmu G_\nu -i 4 G_\lm G_\mu G_\nu]\right\}
\nonumber\\
&=&
\theta_q\, \frac{1}{8\pi^2}\, \ep_{\kp\lm\mu\nu}\, \partial_\kp \tr_{\rm c}
[G_\lm\dmu G_\nu -i (2/3)\,G_\lm G_\mu G_\nu]
=
\theta_q\, \partial_\kp j_\kp^{\rm CS},
\eea
with $j^{\rm CS}_\kp$ the Chern-Simons current
($\tr_{\rm c}$ is the trace in color space). The terms involving
only the $SU(2)$ gauge field are given by
\be
i\intx (n_{\rm c} \theta_q + \theta_\ell) \ep_{\kp\lm\mu\nu}
\frac{1}{192\pi^2}
\partial_\kp \tr_{\rm i}\left(W_\lm\dmu W_\nu - i\half W_\lm W_\mu W_\nu\right)
\ee
where $n_{\rm c}=3$ is the number of colors and $\tr_{\rm i}$ is
the trace in isospin space. This expression has been derived in
the unitary gauge and the  integrand is not explicitly
gauge-invariant anymore. The above expression is furthermore not a
topological object like the divergence of a Chern-Simons current
and being a total derivative it has presumably no physical
significance. Even if it did have the form of a $\theta$ term, it
would still not lead to observable effects according to
\cite{Anselm:1993yz,Anselm:1994uj}. Since we are working in
unitary gauge it is natural to express the remaining contribution
in terms the Z-field, the charged W-fields and the photon field
$A_{\mu}$.
The photon-field contribution is given by
\be
i\intx
\left[n_{\rm c} \left(\frac{4}{9} + \frac{1}{9}\right)
\theta_q +\theta_\ell\right]
\frac{1}{64\pi^2}\,
\ep_{\kp\lm\mu\nu}\,
A_{\kp\lm} A_{\mu\nu},
\ee
The integrand has topological significance in a finite
four-dimensional torus \cite{Smit:1986fn}, but unlike the QCD case
it probably has no physical significance.\footnote{For example,
the corresponding topological suceptibility would scale to zero
like $e^4/\mbox{volume}$ in the infinite-volume limit.} The
remaining terms involving also the W and Z fields are cumbersome
and not particularly illuminating.  As integrals of total
derivatives involving massive fields they are expected to be
physically irrelevant.

\section{Conclusion}
\label{concl}

Using Salcedo's results for the effective action we have shown
that the CP violation in the Standard Model coming from the CKM
matrix is absent to fourth order in the gauge-covariant
derivative-expansion. Six or more orders in the covariant
derivatives of the fields are needed for CKM-type CP violation.
The same holds for the analog mixing matrix in the lepton sector,
which becomes relevant upon extending the SM with Yukawa couplings
such that the neutrinos are given Dirac mass terms. The
possibility of Majorana mass terms in the neutrino sector is very
interesting in the present context as their presence limits the
rephasing invariance, perhaps allowing for a non-zero CP violating
contribution to the effective action already at fourth order. We
leave this question for future investigation.

With a trick of introducing axion-like fields we were able to
recover the known CP-violating total-derivative QCD $\theta$-term,
and electroweak analogs which are not expected to have physical
consequences (see also
\cite{Anselm:1993yz,Anselm:1994uj}).

Last, but not least, the homogeneity of the coefficient functions
calculated by Salcedo strongly suggests that we should not include
the tiny ($\approx 10^{-17}$) product of Yukawa couplings (cf.\
(\ref{dlcpest2})) in order-of-magnitude estimates of CKM
CP-violation at zero temperature. This argument does not apply to
the high-temperature case, for which (\ref{dlcpest1}) may still be
of value.\footnote{Replacing $T$ by the QCD thermal quark mass
at temperatures above the electroweak scale we would gain a factor
$(g_{\rm s}/\sqrt{6})^{-12}\gtrsim 10^{4}$.}

\acknowledgments
I would like to thank Anders Tranberg for useful discussions. This
work received support from FOM/NWO.

\appendix
\section{The functions $N_{123}$ and $N_{1234}$}
\label{appA}
Salcedo's function $N_{123}$ is given by \cite{Salcedo:2000hx}
\begin{equation}
N_{123}= N^R_{123} +N^L_{123}
\log(\sm_1^2/\sm_2^2) - N^L_{321} \log(\sm_3^2/\sm_2^2) \,,
\end{equation}
with
\begin{eqnarray}
N^R_{123} &=&
\frac{1}
{2\sm_1\sm_2\sm_3(\sm_1^2 - \sm_2^2)(\sm_3^2 - \sm_2^2)(\sm_1 - \sm_3)}
\nonumber \\ && \times
\Big(    3\sm_1^2\sm_3^2(\sm_1 - \sm_3)^2
 +  4\sm_1\sm_2\sm_3(\sm_1 + \sm_3)(2\sm_1^2 - 3\sm_1\sm_3 + 2\sm_3^2-\sm_2^2)
\nonumber \\ && {\ \ }
 +  \sm_2^2(\sm_1^4 + 10\sm_1^3\sm_3 - 18\sm_1^2\sm_3^2 +
     10\sm_1\sm_3^3 + \sm_3^4)
 -  \sm_2^4(\sm_1 + \sm_3)^2      \Big)\,,
 \\
N^L_{123} &=&
\frac{2}
{ (\sm_1^2 - \sm_2^2)^2(\sm_1^2 - \sm_3^2)(\sm_1 - \sm_3) }
\nonumber \\ && \times
  \Big(
     \sm_1^4(\sm_2 - 2\sm_3) + \sm_1^2(\sm_2^3 + \sm_3^3) +
  \sm_2^2\sm_3^2(\sm_2 + \sm_3)
\nonumber \\ &&  {\ \ }
+   \sm_1^3(\sm_2^2 - 3\sm_2\sm_3 - \sm_3^2) -
\sm_1\sm_2\sm_3(\sm_2^2 - \sm_3^2)
  \Big) \,.
\end{eqnarray}
This function satisfies $N_{123} = N_{\m{1}\m{2}\m{3}}$,
$N_{123} = -N_{321}$, where $N_{\cdots \m{j}\cdots}\equiv N(\sm_j\to -\sm_j)$;
explicitly:
\be
N(\sm_1,\sm_2,\sm_3)=N(-\sm_1,-\sm_2,-\sm_3),
\quad
N(\sm_1,\sm_2,\sm_3)
= -N(\sm_3,\sm_2,\sm_1),
\label{propN3}
\ee
It is furthermore regular at coinciding arguments. The functions
$N^{(\cdot)}_{123}$ introduced in section \ref{salc} are found by taking
appropriate even-odd combinations. For example, $N^{(12)}_{123}$ is
defined by
\be
N^{(12)}_{123} = \quart\left[
N(\sm_1,\sm_2,\sm_3) - N(-\sm_1,\sm_2,\sm_3) - N(\sm_1,-\sm_2,\sm_3)
+ N(-\sm_1,-\sm_2,\sm_3)\right],
\label{defN12}
\ee
and similar for $N^{(23)}_{123}$ and $N^{(13)}_{123}$,
whereas
\bea
N^{(0)}_{123} &=& \frac{1}{8}\left[
N(\sm_1,\sm_2,\sm_3) + N(-\sm_1,\sm_2,\sm_3) + N(\sm_1,-\sm_2,\sm_3)
+ N(-\sm_1,-\sm_2,\sm_3)
\right.\label{defN0}\\&&\left.\mbox{} +
N(\sm_1,\sm_2,-\sm_3) + N(-\sm_1,\sm_2,-\sm_3) + N(\sm_1,-\sm_2,-\sm_3)
+ N(-\sm_1,-\sm_2,-\sm_3)
\right]
\nonumber
\eea
It follows that $N^{(0)}$ and $f^{(pq)} = N^{(pq)}/\sm_p\sm_q$
are functions of $\sm_1^2$, $\sm_2^2$ and $\sm_3^2$.
The function $N_{1234}$ is given by \cite{Salcedo:2000hx}
\begin{equation}
N_{1234}= N^R_{1234} +N^L_{1234} \log(\sm_1^2) +N^L_{234\m{1}}
\log(\sm_2^2) +N^L_{34\m{1}\m{2}} \log(\sm_3^2)
+N^L_{4\m{1}\m{2}\m{3}} \log(\sm_4^2)\,,
\end{equation}
where
\begin{eqnarray}
N^R_{1234} &=& \frac{1}{4}\Big(
     \frac{2(2\sm_2 + \sm_3)}{
        (\sm_1^2 - \sm_2^2)(\sm_2^2 - \sm_3^2)(\sm_2 - \sm_4)}
 -    \frac{ 2(2\sm_2 + \sm_1 )}{
(\sm_1^2 - \sm_2^2)(\sm_2^2 - \sm_3^2)(\sm_2 +  \sm_4)}
\nonumber \\ &&
-   \frac{3( \sm_2\sm_3 - \sm_1(\sm_2 + \sm_3)) }{
        \sm_3(\sm_1^2 - \sm_3^2)(\sm_2^2 - \sm_3^2)(\sm_3 - \sm_4)}
+  \frac{3(\sm_1\sm_2 - \sm_3(\sm_1 + \sm_2))}{
        \sm_1(\sm_1^2 - \sm_2^2)(\sm_1^2 - \sm_3^2)(\sm_1 + \sm_4)}
\nonumber \\ &&
- \frac{\sm_2\sm_3 + \sm_1(\sm_2 + \sm_3)}{
        \sm_1(\sm_1^2 - \sm_2^2)(\sm_1^2 - \sm_3^2)(\sm_1 - \sm_4)}
 +  \frac{\sm_2\sm_3 + \sm_1(\sm_2 + \sm_3)}{
        \sm_3(\sm_1^2 - \sm_3^2)(\sm_2^2 - \sm_3^2)(\sm_3 + \sm_4)}
\nonumber \\ &&
+   \frac{1}{\sm_1\sm_2\sm_3\sm_4}
\Big)\,,
\end{eqnarray}
\begin{eqnarray}
N^L_{1234} &=&
\frac{1}{
     2(\sm_1^2 - \sm_2^2)^2(\sm_1^2 - \sm_3^2)^2(\sm_1^2 - \sm_4^2)^2}
\nonumber \\ && \times
\Big(
     6\sm_1^7\sm_3 + (\sm_2 - \sm_4) (\sm_2^2\sm_3^3\sm_4^2+3\sm_1^6\sm_3)
- \sm_1\sm_2\sm_3^3\sm_4(\sm_2 - \sm_4)^2 +
\nonumber \\ &&
 +  \sm_1^2\sm_3^2(\sm_2^3(2\sm_4 + \sm_3 ) - \sm_4^3(2\sm_2 + \sm_3))
\nonumber \\ &&
-    \sm_1^4(\sm_2 - \sm_4)(2\sm_2^2(\sm_3 + \sm_4) +
     \sm_2\sm_4(\sm_3 + 2\sm_4)
+  2\sm_3(\sm_3^2 + \sm_4^2))
\nonumber \\ &&
+ \sm_1^3
        (-\sm_3^2\sm_4^3 + \sm_2^2\sm_4^2(2\sm_3 + \sm_4) +
        \sm_2\sm_3\sm_4(2\sm_3^2 + \sm_4^2)
\\&&
        + \sm_2^3(-\sm_3^2 + \sm_3\sm_4 + \sm_4^2))
\nonumber \\ &&
    -   \sm_1^5(\sm_2^2(4\sm_3 + \sm_4)
+ \sm_2(-\sm_3^2 + 2\sm_3\sm_4 + \sm_4^2) +
\sm_3(2\sm_3^2 - \sm_3\sm_4 + 4\sm_4^2))\Big).
\nonumber
\end{eqnarray}
It is regular at coinciding arguments and has the symmetries
\be
N_{1234} = N_{\m{1}\m{2}\m{3}\m{4}},
\quad
N_{1234} = N_{234\m{1}},
\quad
N_{1234} = - N_{4321}.
\label{propN4}
\ee
Similar to the case of $N_{123}$ we can define
$N^{(0)}_{1234}$, $N^{(12)}_{1234}$, \ldots,
$N^{(34)}_{1234}$, $N^{(1234)}_{1234}$, such that $N^{(0)}_{1234}$,
$f^{(pq)}_{1234} = N^{(pq)}_{1234}/\sm_p\sm_q$, and
$f^{(1234)}_{1234}=N^{(1234)}_{1234}/\sm_1\sm_2\sm_4\sm_4$
are functions of $\sm_1^2$, \ldots, $\sm_4^2$, with
\be
N_{1234} = N^{(0)}_{1234}+ \sum_{p<q} f^{(pq)}_{1234}\sm_p\sm_q
+ f^{(1234)}_{1234}\sm_1\sm_2\sm_3\sm_4.
\ee

\section{$\Gm_{\rm gWZW}$}
\label{appB}

The gauged Wess-Zumino-Witten action is given in
\cite{Salcedo:2000hx} using the notation of differential forms, in
addition to the earlier used notationl conventions in section
\ref{salc}. The following one-forms are introduced
\cite{Salcedo:2000hx}:
\bea
\sR &=& \sm^{-1}d\sm,
\quad \sL = \sm d\sm^{-1},
\\
\sR_c &=& \sm^{-1}\sDa\sm = \sR+\sm^{-1}\sv\sm - \sv,
\\
\sL_c &=& \sm\sDa\sm^{-1} = -\sDa\sm\sm^{-1}  = \sL+\sm\sv\sm^{-1} - \sv =
-\sm\sR_c\sm^{-1}.
\eea
Here $\sR_c$ and $\sL_c$ are covariant under $U(n)\times U(n)$
gauge transformations. The extended gauged WZW action is given by
\bea
\Gamma_{\rm gWZW}[v,m]
&=&  \frac{1}{48\pi^2} \int \tr \left[
-\frac{1}{5}\sR_c^5
+ (\sR_c^3+\sL_c^3)\sF
- 2(\sR_c+\sL_c)\sF^2
\right. \nonumber \\ && \left.
-\sR_c\sF\sm^{-1}\sF\sm -\sL_c\sF\sm\sF\sm^{-1}
 -4\sv\sF^2  +2\sv^3\sF  -\frac{2}{5}\sv^5
\label{gWZW1}
 \right].
\eea
The integral is over a five-dimensional manifold with the physical
four-dimensional space-time as boundary (the fields have been extended into
a fifth dimension, $\sv_5$ can be taken equal to zero). Most of the
integrations over the fifth dimension can be done, except for a term
involving the WZW five-form $\sR^5$, and an equivalent expression
\cite{Salcedo:2000hx} for a $\Gamma_{\rm gWZW}$ is given in (\ref{gWZW2}).

The last three terms in (\ref{gWZW1}) are not gauge invariant,
they correspond to the $U(n)\times U(n)$ chiral anomaly. Their
reduction to the Standard Model should be gauge invariant, since
the SM is anomaly free 
\cite{Bouchiat:1972iq,Gross:1972pv,D'Hoker:1984ph}, 
which can be seen as
follows. The potentially non-invariant terms constitute the
Chern-Simons form, with exterior derivative
\be
d\,\tr\left(-4\sv\sF^2 + 2\sv^3\sF - \frac{2}{5}\, \sv^5\right)
=
-4\,\tr\,\sF^3.
\ee
Introducing a six-dimensional manifold with the 5D manifold as boundary, and extending the gauge field into this 6D domain, we may write
\be
\int \tr\left(-4\sv\sF^2 + 2\sv^3\sF - \frac{2}{5}\, \sv^5\right)
=
-4 \int \tr\,\sF^3.
\ee
Writing the gauge field in terms of its generators,
$F_{\mu\nu} = F_{\mu\nu}^p T_p$, we have
\be
\tr\, \sF^3 = {\rm str}(T_p T_q T_r)\,
\sF_{\kp\lm}^p\sF_{\mu\nu}^q\sF_{\rh\sg}^r\,
dx^{\kp}\wedge dx^{\lm}\cdots\wedge dx^{\sg},
\ee
in which ${\rm str}(T_p T_q T_r)$ is the symmetrized trace (since
only the part of the trace that is symmetric under permutations of
$p$, $q$ and $r$ contributes). For an anomaly-free representation
of the gauge group
${\rm str}(T_p T_q T_r)=0$, and for the reduction to the Standard
Model $\Gm_{\rm gWZW}$ is gauge invariant.

\bibliography{lit}
\end{document}

%% file: newcomm.tex
\newcommand{\intx}{\int d^4 x\,}

\newcommand{\gm}{\gamma}
\newcommand{\dl}{\delta}
\newcommand{\ep}{\epsilon}

\newcommand{\et}{\eta}
\newcommand{\kp}{\kappa}
\newcommand{\lm}{\lambda}
\newcommand{\rh}{\rho}
\newcommand{\sg}{\sigma}
\newcommand{\ta}{\tau}

\newcommand{\vr}{\varphi}

\newcommand{\ps}{\psi}
\newcommand{\om}{\omega}
\newcommand{\Ps}{\Psi}

\newcommand{\Psb}{\bar{\Psi}}
\newcommand{\Ph}{\Phi}
\newcommand{\Phd}{\Phi^{\dagger}}
\newcommand{\Om}{\Omega}

\newcommand{\Gm}{\Gamma}
\newcommand{\Lm}{\Lambda}

\newcommand{\half}{\frac{1}{2}}
\newcommand{\quart}{\frac{1}{4}}

\newcommand{\Tr}{\mbox{Tr}\,}  
\newcommand{\tr}{\mbox{tr}\,}  
\newcommand{\unity}{1\!\! 1}

\newcommand{\dmu}{\partial_{\mu}}

\newcommand{\dnu}{\partial_{\nu}}

\newcommand{\psb}{\bar{\psi}}

\newcommand{\eela}[1]{\label{#1}\end{equation}}
\newcommand{\eeala}[1]{\label{#1}\end{eqnarray}}
\newcommand{\be}{\begin{equation}}
\newcommand{\ee}{\end{equation}}
\newcommand{\bea}{\begin{eqnarray}}
\newcommand{\eea}{\end{eqnarray}}

%% file: effcp.bbl
\providecommand{\href}[2]{#2}\begingroup\raggedright\begin{thebibliography}{10}

\bibitem{KT}
E.~W. Kolb and M.~S. Turner, {\em The {E}arly {U}niverse}.
\newblock Addison-Wesley, Reading, {M}assachusetts, 1990.

\bibitem{Cabibbo:1963yz}
N.~Cabibbo, {\it Unitary symmetry and leptonic decays},  {\em Phys. Rev. Lett.}
  {\bf 10} (1963) 531--532.

\bibitem{Kobayashi:1973fv}
M.~Kobayashi and T.~Maskawa, {\it {CP} violation in the renormalizable theory
  of weak interaction},  {\em Prog. Theor. Phys.} {\bf 49} (1973) 652--657.

\bibitem{Eidelman}
{\bf Particle Data Group} Collaboration, S.~Eidelman {\em et.~al.}, {\it Review
  of particle physics},  {\em Phys. Lett.} {\bf B592} (2004) 1.

\bibitem{Kuzmin:1985mm}
V.~A. Kuzmin, V.~A. Rubakov, and M.~E. Shaposhnikov, {\it On the anomalous
  electroweak baryon number nonconservation in the early universe},  {\em Phys.
  Lett.} {\bf B155} (1985) 36.

\bibitem{Cohen:1993nk}
A.~G. Cohen, D.~B. Kaplan, and A.~E. Nelson, {\it Progress in electroweak
  baryogenesis},  {\em Ann. Rev. Nucl. Part. Sci.} {\bf 43} (1993) 27--70,
  [\href{http://xxx.lanl.gov/abs/hep-ph/9302210}{{\tt hep-ph/9302210}}].

\bibitem{Rubakov:1996vz}
V.~A. Rubakov and M.~E. Shaposhnikov, {\it Electroweak baryon number
  non-conservation in the early universe and in high-energy collisions},  {\em
  Usp. Fiz. Nauk} {\bf 166} (1996) 493--537,
  [\href{http://xxx.lanl.gov/abs/hep-ph/9603208}{{\tt hep-ph/9603208}}].

\bibitem{Shaposhnikov:1987tw}
M.~E. Shaposhnikov, {\it Baryon asymmetry of the universe in standard
  electroweak theory},  {\em Nucl. Phys.} {\bf B287} (1987) 757--775.

\bibitem{Shaposhnikov:1988pf}
M.~E. Shaposhnikov, {\it Structure of the high temperature gauge ground state
  and electroweak production of the baryon asymmetry},  {\em Nucl. Phys.} {\bf
  B299} (1988) 797.

\bibitem{Jarlskog:1985ht}
C.~Jarlskog, {\it Commutator of the quark mass matrices in the standard
  electroweak model and a measure of maximal {CP} violation},  {\em Phys. Rev.
  Lett.} {\bf 55} (1985) 1039.

\bibitem{BigiSanda}
I.~Bigi and A.~Sanda, {\em C{P} {V}iolation}.
\newblock Cambridge University Press, Cambridge, {UK}, 1999.

\bibitem{Branco:1999fs}
G.~C. Branco, L.~Lavoura, and J.~P. Silva, {\it {CP V}iolation}, . Oxford, UK:
  Clarendon (1999).

\bibitem{Garcia-Bellido:1999sv}
J.~Garc\'{\i}a-Bellido, D.~Y. Grigoriev, A.~Kusenko, and M.~E. Shaposhnikov,
  {\it Non-equilibrium electroweak baryogenesis from preheating after
  inflation},  {\em Phys. Rev.} {\bf D60} (1999) 123504,
  [\href{http://xxx.lanl.gov/abs/hep-ph/9902449}{{\tt hep-ph/9902449}}].

\bibitem{Krauss:1999ng}
L.~M. Krauss and M.~Trodden, {\it Baryogenesis below the electroweak scale},
  {\em Phys. Rev. Lett.} {\bf 83} (1999) 1502--1505,
  [\href{http://xxx.lanl.gov/abs/hep-ph/9902420}{{\tt hep-ph/9902420}}].

\bibitem{Garcia-Bellido:2003wd}
J.~Garc\'{\i}a-Bellido, M.~Garc\'{\i}a-P\'erez, and A.~Gonz\'alez-Arroyo, {\it
  Chern-{S}imons production during preheating in hybrid inflation models},
  {\em Phys. Rev.} {\bf D69} (2004) 023504,
  [\href{http://xxx.lanl.gov/abs/hep-ph/0304285}{{\tt hep-ph/0304285}}].

\bibitem{Tranberg:2003gi}
A.~Tranberg and J.~Smit, {\it Baryon asymmetry from electroweak tachyonic
  preheating},  {\em JHEP} {\bf 11} (2003) 016,
  [\href{http://xxx.lanl.gov/abs/hep-ph/0310342}{{\tt hep-ph/0310342}}].

\bibitem{Copeland:2001qw}
E.~J. Copeland, D.~Lyth, A.~Rajantie, and M.~Trodden, {\it Hybrid inflation and
  baryogenesis at the {TeV} scale},  {\em Phys. Rev.} {\bf D64} (2001) 043506,
  [\href{http://xxx.lanl.gov/abs/hep-ph/0103231}{{\tt hep-ph/0103231}}].

\bibitem{vanTent:2004rc}
B.~van Tent, J.~Smit, and A.~Tranberg, {\it Electroweak-scale inflation,
  inflaton-{H}iggs mixing and the scalar spectral index},
  \href{http://xxx.lanl.gov/abs/hep-ph/0404128}{{\tt hep-ph/0404128}}.

\bibitem{Skullerud:2003ki}
J.-I. Skullerud, J.~Smit, and A.~Tranberg, {\it W and {H}iggs particle
  distributions during electroweak tachyonic preheating},  {\em JHEP} {\bf 08}
  (2003) 045, [\href{http://xxx.lanl.gov/abs/hep-ph/0307094}{{\tt
  hep-ph/0307094}}].

\bibitem{Garcia-Bellido:2002aj}
J.~Garc\'{\i}a-Bellido, M.~Garc\'{\i}a-P\'erez, and A.~Gonz\'alez-Arroyo, {\it
  Symmetry breaking and false vacuum decay after hybrid inflation},  {\em Phys.
  Rev.} {\bf D67} (2003) 103501,
  [\href{http://xxx.lanl.gov/abs/hep-ph/0208228}{{\tt hep-ph/0208228}}].

\bibitem{Farrar:1993sp}
G.~R. Farrar and M.~E. Shaposhnikov, {\it Baryon asymmetry of the universe in
  the minimal {S}tandard {M}odel},  {\em Phys. Rev. Lett.} {\bf 70} (1993)
  2833--2836, [\href{http://xxx.lanl.gov/abs/hep-ph/9305274}{{\tt
  hep-ph/9305274}}]. [Erratum-ibid.71:210,1993].

\bibitem{Farrar:1994hn}
G.~R. Farrar and M.~E. Shaposhnikov, {\it Baryon asymmetry of the universe in
  the standard electroweak theory},  {\em Phys. Rev.} {\bf D50} (1994) 774,
  [\href{http://xxx.lanl.gov/abs/hep-ph/9305275}{{\tt hep-ph/9305275}}].
  [Erratum-ibid.71:210,1993].

\bibitem{Farrar:1994kf}
G.~R. Farrar and M.~E. Shaposhnikov, {\it Note added to '{B}aryon asymmetry of
  the universe in the standard model'},
  \href{http://xxx.lanl.gov/abs/hep-ph/9406387}{{\tt hep-ph/9406387}}.

\bibitem{Konstandin:2003dx}
T.~Konstandin, T.~Prokopec, and M.~G. Schmidt, {\it Axial currents from {CKM}
  matrix {CP} violation and electroweak baryogenesis},  {\em Nucl. Phys.} {\bf
  B679} (2004) 246--260, [\href{http://xxx.lanl.gov/abs/hep-ph/0309291}{{\tt
  hep-ph/0309291}}].

\bibitem{Salcedo:2000hp}
L.~L. Salcedo, {\it Derivative expansion for the effective action of chiral
  gauge fermions: The normal parity component},  {\em Eur. Phys. J.} {\bf C20}
  (2001) 147--159, [\href{http://xxx.lanl.gov/abs/hep-th/0012166}{{\tt
  hep-th/0012166}}].

\bibitem{Salcedo:2000hx}
L.~L. Salcedo, {\it Derivative expansion for the effective action of chiral
  gauge fermions: The abnormal parity component},  {\em Eur. Phys. J.} {\bf
  C20} (2001) 161--184, [\href{http://xxx.lanl.gov/abs/hep-th/0012174}{{\tt
  hep-th/0012174}}].

\bibitem{Bouchiat:1972iq}
C.~Bouchiat, J.~Iliopoulos, and P.~Meyer, {\it An anomaly free version of
  {W}einberg's model},  {\em Phys. Lett.} {\bf B38} (1972) 519--523.

\bibitem{Gross:1972pv}
D.~J. Gross and R.~Jackiw, {\it Effect of anomalies on quasirenormalizable
  theories},  {\em Phys. Rev.} {\bf D6} (1972) 477--493.

\bibitem{D'Hoker:1984ph}
E.~D'Hoker and E.~Farhi, {\it Decoupling a fermion whose mass is generated by a
  {Y}ukawa coupling: The general case},  {\em Nucl. Phys.} {\bf B248} (1984)
  59.

\bibitem{LeBellac96}
M.~Le~Bellac, {\em Thermal Field Theory}.
\newblock Cambridge University Press, Cambridge, UK, 1996.

\bibitem{Anselm:1993yz}
A.~A. Anselm and A.~A. Johansen, {\it Baryon nonconservation in standard model
  and {Y}ukawa interaction},  {\em Nucl. Phys.} {\bf B407} (1993) 313--330.

\bibitem{Anselm:1994uj}
A.~A. Anselm and A.~A. Johansen, {\it Can electroweak theta term be
  observable?},  {\em Nucl. Phys.} {\bf B412} (1994) 553--573,
  [\href{http://xxx.lanl.gov/abs/hep-ph/9305271}{{\tt hep-ph/9305271}}].

\bibitem{Smit:1986fn}
J.~Smit and J.~C. Vink, {\it Remnants of the index theorem on the lattice},
  {\em Nucl. Phys.} {\bf B286} (1987) 485.

\end{thebibliography}\endgroup
